\documentclass[prl,aps,twocolumn,groupedaddress,floats,showpacs]{revtex4}
\usepackage{graphicx}
\usepackage{dcolumn}
\usepackage{bm}
\usepackage{color}
\usepackage{mathtools}
\begin{document}

\newcounter{fnnumber}
%----felix shorhand: -------
\newcommand{\bea}{\begin{eqnarray}}
\newcommand{\eea}{\end{eqnarray}}
\newcommand{\be}{\begin{equation}}
\newcommand{\ee}{\end{equation}}
\newcommand{\bes}{\begin{equation*}}
\newcommand{\ees}{\end{equation*}}
\newcommand{\ds}{\displaystyle}
\newcommand{\rr}{\mathbf{r}}
\newcommand{\kk}{\mathbf{k}}
\newcommand{\pp}{\mathbf{p}}
\newcommand{\qq}{\mathbf{q}}
\newcommand{\ra}{\rangle}
\newcommand{\la}{\langle}
\newcommand{\si}{\sigma}
\newcommand{\sip}{{\sigma'}}
\newcommand{\up}{\uparrow}
\newcommand{\down}{\downarrow}
\newcommand{\GtN}{\tilde{G}_{\mathcal{N}}}
\newcommand{\Gt}{\tilde{G}}
\newcommand{\Nr}{\mathcal{N}}
\newcommand{\Dr}{\mathcal{D}}
\newcommand{\Cr}{\mathcal{C}}
\newcommand{\bL}{L} %{\mathb{\Lambda}}
\renewcommand{\Pr}{\mathcal{P}}
\newcommand{\Sr}{\mathcal{S}}
\newcommand{\blue}{\color{blue}}
%---------------------------------

%%%%%%%%%%%%%%%%%%%%%%%%%%%%%%%%%%%%%% AUTHORS %%%%%%%%%%%%%%%%%%%%%%%%%
\author{Riccardo Rossi$^1$ and F\'elix Werner$^2$}

\affiliation{Laboratoire de Physique Statistique$\,^1$ and Laboratoire Kastler Brossel$\,^2$, Ecole Normale Sup\'erieure, 
UPMC, CNRS,
Universit\'e Paris Diderot$\,^1$,
Coll\`ege de France$\,^2$,
24 rue Lhomond,
 75005 Paris, France}
 
 \author{Nikolay Prokof'ev$^{3,4}$ and Boris Svistunov$^{3,4,5}$}
\affiliation{$^3$Department of Physics, University of Massachusetts,
Amherst, MA 01003, USA} 
\affiliation{$^4$Russian Research Center
``Kurchatov Institute,'' 123182 Moscow, Russia}

\affiliation{$^5$Wilczek Quantum Center, Zhejiang University of Technology, Hangzhou 310014, China}

%%%%%%%%%%%%%%%%%%%%%%%%%%%%%%%%%%%%%%%%%%%%%%%%%%%%%%%%%%%%%%%%%%%%%%%%%%%%%%

\title{
  Shifted-Action Expansion and
  Applicability of Dressed Diagrammatic Schemes
}

%%%%%%%%%%%%%%%%%%%%%%%%%%%%%%%%%%%%%%%%%%%%%%%%%%%%%%%%%%%%%%%%%%%%%%%%%%%%%%
\begin{abstract}

%  Alors que les series diagrammatiques bare sont des simples developpements de Taylor en l'interaction,  les series diagrammatiques squelettes construites sur des propagateurs et/ou vertex habilles (partiellement ou completement) sont habituellement construites en reordonnant une serie bare, ce qui est une manipulation a priori injustifiee.  Ceci peut effectivement conduire a une convergence vers un resultat faux, comme montre recemment [Kozik].  Ici nous montrons que pour une broad classe de schemas diagrammatiques habilles,  il existe une action dependante d'un parametre auxiliaire $\xi$ telle que le developpement de Taylor en $\xi$ reproduise  la serie diagrammatique squelette.

  While bare diagrammatic series are merely Taylor expansions in powers of interaction strength,
 dressed diagrammatic series, built on fully or partially dressed lines and vertices,
  are usually constructed by reordering the bare diagrams, which is an {\it a priori} unjustified manipulation,
  and can even lead to convergence to an unphysical result [Kozik, Ferrero, and Georges,
%    Phys. Rev. Lett. {\bf 114}, 156402 (2015)].
    PRL~{\bf 114},~156402~(2015)].
  Here we show that for a broad class of partially dressed diagrammatic schemes, there exists an action~$S^{(\xi)}$ depending analytically
  on an auxiliary complex parameter $\xi$,
  such that the Taylor expansion in $\xi$ of correlation functions reproduces the original diagrammatic series.
  The resulting  applicability conditions are similar
 % controllability
  to the bare case.
%  [The auxiliary action framework is very general and includes most partially   dressed diagrammatic schemes as particular cases (?)].
  For fully dressed skeleton diagrammatics,
analyticity of $S^{(\xi)}$ is not granted,
%this
%leads us to
and we
formulate a
sufficient condition for converging to the correct
 result.
%answer.
%Furthermore, we propose a partially dressed scheme which is free of such convergence issues.

\end{abstract}

% 71.10.Fd	Lattice fermion models (Hubbard model, etc.)
% 02.70.Ss	Quantum Monte Carlo methods

\maketitle

Much of theoretical physics is formulated in the language of Feynman diagrams,
%built on partially or fully dressed propagators or vertices.
%Feynman diagrams are the language of theoretical physics
in various fields such as
condensed matter,
nuclear physics,
%statistical physics,
and QCD.
A powerful feature of the diagrammatic technique, used in each of the above fields, is
 the possibility to build
 diagrams  on
 partially or fully dressed propagators or vertices, see, {\it e.g.}, Refs.~\cite{PitaevskiiLifchitz,AGD,RevuePhysNuclSelfconsDiag,revue_qcd_resummation_hot,revue_qcd_color_SC}.
In quantum many-body physics, notable examples include dilute gases,
whose description is radically improved 
 if ladder diagrams are summed up so that the
 expansion is done in terms of the scattering amplitude instead of the bare interaction potential,
 and Coulomb interactions,
 which one has to screen to have a meaningful diagrammatic technique.

 With the development of Diagrammatic Monte Carlo,
it becomes possible to compute Feynman
diagrammatic expansions to high order for fermionic strongly correlated quantum many-body problems~\cite{VanHoucke1,VanHouckeEPL,VanHouckeEOS,KulaginPRL,MishchenkoProkofevPRL2014,DengEmergentBCS}.
The number of diagrams grows factorially with the order,
even for a fully irreducible skeleton scheme~\cite{molinari_manini_enumeration}.
%Il y a un nombre factoriel de diagrammes.
Nevertheless,
for fermionic systems on a lattice at finite temperature, diagrammatic series
(of the form $\sum_n a_n$ with $a_n$ the sum of all order-$n$ diagrams)
are typically convergent in a broad range of parameters,
due to a nearly perfect cancellation of contributions of different sign within each order,
as proven mathematically~\cite{Mastropietro} and seen numerically~\cite{VanHoucke1,VanHouckeEPL,KulaginPRL,MishchenkoProkofevPRL2014,DengEmergentBCS}.

One might think that partial or full renormalization of diagrammatic elements (propagators,
interactions, vertices, etc.) always leads to more compact and better behaving diagrammatic expansion.
However, such a dressed diagrammatic series cannot be used blindly:
Even when it converges, the result
is not guaranteed to be correct,
since it is {\it a priori} not allowed to reorder the terms of a series that is not absolutely convergent
(the sum of the absolute values of {\it individual diagrams} is typically infinite,
due to factorial scaling of the number of diagrams with the order). And indeed,
for a skeleton series, {\it i.e.}, a series built on the fully dressed propagator,
convergence to a wrong result does occur
in the case
of the Hubbard model in
the strongly correlated regime near half filling~\cite{Kozik_2_solutions},
and
preliminary results suggest that
the corresponding self-consistent skeleton scheme converges to a wrong result
as a function of the maximal self-energy diagram order $\Nr$~\cite{Deng_Kozik_private}.
Both of these phenomena are clearly seen
in the exactly solvable zero space-time dimensional case~\cite{Rossi_Werner_0+0dim,Rossi_Werner_unpub}.

In this work,
we
establish a
condition that is necessarily violated in the event of convergence to a wrong result
of the self-consistent skeleton scheme.
Furthermore,
we show that this convergence issue is absent for a broad class of partially dressed schemes.
In particular, we propose a simple scheme based on the truncated skeleton series.
The underlying idea 
is to construct an action $S^{(\xi)}$ that depends on an auxiliary complex parameter $\xi$ such that the Taylor series in $\xi$ of correlation functions
reproduces the
dressed
diagrammatic  series built on a given partially or fully dressed propagator.
%  The dressed character of the expansion is enforced by harmonic $\xi$-dependent  counter-terms,  which exactly cancel out the reducible diagrams.
  This makes the dressed scheme as mathematically justified as a bare scheme,
  provided
  $S^{(\xi)}$ is analytic with respect to $\xi$ and  $S^{(\xi=1)}$ coincides with the physical action;
  these conditions hold automatically in the partially dressed case,
  while in the fully dressed case they hold under a simple sufficient condition which we provide.
  Our construction applies to a general class of diagrammatic schemes
  built on dressed lines and vertices,
  including two-particle ladders and
  screened long-ranged potentials.
  
%%%%%%%%%%%%%%%%%%%%%%%%%%%%%
%  {\it Boldifying the single-particle propagator.}
{\it Partially dressed single-particle propagator.}
  We consider a generic fermionic many-body problem described by an action
  \begin{equation}
S[\psi,\bar{\psi}]  = \langle \psi | G_0^{-1} | \psi \rangle  + S_{\rm int}[\psi,\bar{\psi}] 
\label{action_phys}
\end{equation}
where $\psi, \bar{\psi}$ are Grassmann fields~\cite{NegeleOrland},
and we use bra-ket notations to suppress space, imaginary time, possible internal quantum numbers,
and integrals/sums over them,
{
  {\it i.e.},
$\la \psi | G_0^{-1} | \psi \ra$ denotes
the integral/sum over $\rr$, $\tau$ and $\sigma$ of
%$\sum_{\rr,\sigma} \int_0^\beta d\tau\,
$\bar{\psi}_\sigma(\rr,\tau)\,(G_{0,\sigma}^{-1}\,\psi_\sigma)(\rr,\tau)$.}
$G_0^{-1}$ stands for the inverse, in the sense of operators, of the free propagator.
%Denoting by $G$ the full propagator,
The full propagator $G$ and the self-energy $\Sigma$ are related
through the Dyson equation
$G^{-1}  = G_0^{-1}-\Sigma$.
The bare Feynman diagrammatic expansion corresponds to
perturbation theory in $S_{\rm int}$.
In order to generate a  diagrammatic expansion
built on a partially dressed single-particle propagator $\tilde{G}_{\cal N}$,
we introduce an auxiliary
action of the form
\begin{equation}
S_\Nr^{(\xi)}[\psi,\bar{\psi}]  =  \langle \psi | G_{0,\Nr}^{-1}(\xi)| \psi \rangle  + \xi S_{\rm int}[\psi,\bar{\psi}],
\label{action_aux}
\end{equation}
where
\be
G_{0,\Nr}^{-1}(\xi) = 
\tilde{G}_{\cal N} ^{-1} \! + \xi \Lambda_1   + \ldots + \xi^{\cal N} \Lambda_{\cal N},
\label{eq:G0_N_xi}
\ee
$\xi$ is an auxiliary complex parameter,
and $\Lambda_1, \dots, \Lambda_\Nr$ are appropriate operators. 
%and loosely make no distinction between operators and the corresponding kernels.
$\tilde{G}_{\cal N}$ is the single particle propagator
for $S^{(\xi = 0)}_\Nr$.
At $\xi \neq 0$, one can still view $\tilde{G}_{\cal N} $ as the free propagator, provided one includes in the interaction terms
{ not only $\xi S_{\rm int}$, but also the quadratic terms $\la \psi | \xi^n \Lambda_n | \psi \ra$.}
Accordingly, $\xi$ is interpreted as a coupling constant,
and the $\xi^n \Lambda_n$ acquire the meaning of counter-terms.
These counter-terms
can be tuned to cancel out reducible diagrams,
thereby enforcing
the dressed character of the diagrammatic expansion.
%with respect to the $n$-th order self-energy generated by the anharmonic interaction term $\xi S_{\rm int}$.
A natural requirement is that
$S_\Nr^{(\xi=1)}$ coincides with the physical action $S$,
{\it i.e.}, that
\begin{equation}
 \tilde{G}_{\cal N}^{-1} + \sum_{n=1}^{\cal N} \Lambda_n   =\,  G_0^{-1}.
\label{G_0_constr}
\end{equation}
For given $G_0$, this should be viewed as an equation to be solved for
$\tilde{G}_{\cal N}$  (it is non-linear if the $\Lambda_n$'s depend on $\tilde{G}_{\cal N}$).
%The $\Lambda_n$'s can be viewed as shifts, since
The unperturbed action for the dressed expansion,
$\la \psi | \GtN^{-1} | \psi\ra$,
is shifted by the $\Lambda_n$'s with respect to the unperturbed action
for the bare expansion,
$\la \psi | G_0^{-1} | \psi\ra$.
%and we assume that the solution exists.

We can then use any action of the generic class (\ref{action_aux}) for producing physical answers in the form of Taylor expansion in powers of $\xi$, provided the propagator $\tilde{G}_\Nr$
and the shifts  $\Lambda_n$ satisfy Eq.~(\ref{G_0_constr}).
More precisely, consider
the full single-particle propagator $G_\Nr(\xi)$  of the action $S^{(\xi)}_\Nr$,
and the corresponding
self-energy 
\be
\Sigma_\Nr(\xi) \coloneqq G_{0,\Nr}^{-1}(\xi) - G_\Nr^{-1}(\xi).
\label{eq:Dyson_N_xi}
\ee
{ Note that since $S_\Nr^{(\xi=1)}=S$, we have $G_\Nr(\xi{=}1) = G$ and hence also $\Sigma_\Nr(\xi{=}1)=\Sigma$.}
We assume for simplicity that $\Sigma_\Nr(\xi)$ is analytic at $\xi=0$,
and that { its Taylor series
  %which we shall denote by
  $\sum_{n=1}^\infty \Sigma_\Nr^{(n)}[\GtN]\,\xi^n$},
%\be
%\Sigma_\Nr(\xi) = \sum_{n=1}^\infty \Sigma_\Nr^{(n)}[\GtN]\,\xi^n
%\label{eq:Taylor_Sigma_Nr}
%\ee
converges
at $\xi=1$.
We expect these assumptions to hold for fermionic lattice models at finite temperature in a broad parameter regime,
given that the action $S^{(\xi)}_\Nr$ is analytic in $\xi$~\cite{Mastropietro,rivasseau_tree_expansion,VanHoucke1,VanHouckeEPL,KulaginPRL,MishchenkoProkofevPRL2014,DengEmergentBCS}.
  Then,
since $S^{(\xi)}_\Nr$ is an entire function of $\xi$,
% by analytic continuation,
 %\footnote{ This reasoning is expected to hold even in the event of crossing a first order phase transition as a function of $\xi$,  since the series should then diverge at $\xi=1$~\cite{Isakov}.}
 we can conclude that
 \be
 \Sigma =\sum_{n=1}^\infty \Sigma_\Nr^{(n)}[\GtN],
 \label{eq:Sig=Diag}
 \ee
{\it i.e.}, the physical self-energy
%$\Sigma = \Sigma_\Nr(\xi=1)$
is equal to
the dressed diagrammatic series.

{ This last step of the reasoning can be justified using} the following presumption:
{\it Let $\Dr$ be a connected open region of the complex plane containing $0$. Assume that $S^{(\xi)}$ is analytic in $\Dr$, that the corresponding self-energy $\Sigma(\xi)$ is analytic at $\xi=0$, and that $\Sigma(\xi)$ admits an analytic continuation $\tilde{\Sigma}(\xi)$ in $\Dr$. Then, $\Sigma$ and $\tilde{\Sigma}$ coincide on $\Dr$.}
This presumption
is based on the following argument:
Since $S^{(\xi)}$ is analytical,
if no phase transition occurs when varying $\xi$ in $\Dr$, then 
$\Sigma(\xi)$ is analytical on $\Dr$, and by the identity theorem for analytic functions, $\Sigma$ and $\tilde{\Sigma}$ coincide on $\Dr$.
If a phase transition would be crossed as a function of $\xi$ in $\Dr$, 
analytic continuation
through the phase transition would not be possible~\cite{Isakov},
contradicting the above assumption on the existence of $\tilde{\Sigma}$.
%In this article we restrict to convergent diagrammatic series, in which case $\Dr$ is a disc.)
{ 
Applying this presumption to $\tilde{\Sigma}(\xi) := \sum_{n=1}^\infty \Sigma_\Nr^{(n)}[\tilde{G}_\Nr]\,\xi^n$, which has a radius of congergence $R\geq1$
(from the Cauchy-Hadamard theorem),
and taking for $\Dr$ the open disc of radius $R$,
we directly obtain Eq.~(\ref{eq:Sig=Diag}) provided $R>1$.
If $R{=}1$, we can still derive Eq.~(\ref{eq:Sig=Diag}),  using Abel's theorem and assuming that $\Sigma_\Nr(\xi)$ is continuous at $\xi{=}1$,
which, given that the action is entire in $\xi$, is generically expected (except for physical parameters fined-tuned precisely to a first-order phase transition, where $\Sigma$ is not uniquely defined).}

\setcounter{fnnumber}{\thefootnote}

  %;
%however, close to the transition, the series may appear to converge in practice, yielding the properties of the metastable phase; the broken symmetry phase may then be accessed through a separate computation using, {\it e.g.}, anomalous propagators.}.

{\it Semi-bold scheme.}
We first focus on the choice
\be
\Lambda_n = \Sigma_{\rm bold}^{(n)}[\GtN]\ \ \ \ \ (1\leq n \leq \Nr),
\label{eq:Lambda_semi_b}
\ee
where $\Sigma_{\rm bold}^{(n)}[\mathcal{G}]$
is the sum of all skeleton diagrams of order $n$, built with the propagator $\mathcal{G}$
{ and the bare interaction vertex corresponding to $S_{\rm int}$,}
that remain connected when cutting two $\mathcal{G}$ lines.
This means that $\GtN$ is the solution of the bold scheme for maximal order $\Nr$, cf.~Eq.~(\ref{G_0_constr}).
%We will refer to this choice of $\GtN$ as `skeleton sequence'.
For a given $\Nr$,
higher-order dressed graphs can then be built on
$\GtN$. 
%as the partially dressed propagator to generate higher-order self-energy diagrams
The numerical protocol
corresponding to this `semi-bold' scheme
consists of
two independent parts: %\\
Part I is the Bold Diagrammatic Monte Carlo simulation of the truncated order-${\cal N}$ skeleton sum
employed to solve iteratively for $ \tilde{G}_{\cal N}$ satisfying Eqs.~(\ref{G_0_constr},\ref{eq:Lambda_semi_b}); %\\
Part II is the diagrammatic Monte Carlo simulation of higher-order terms,
$\Sigma_\Nr^{(n)}[\GtN], n>\Nr$,
that uses
$ \tilde{G}_{\cal N}$ as the bare propagator.
%The bare expansion is ``renormalized" in the sense that all self-energy-type insertions of order ${\cal N}$ are absent since, by construction, they are compensated by counter-terms in the auxiliary action (\ref{action_aux}).
%\\
{ Note that here $\Nr$ is fixed (contrarily to the conventional skeleton scheme discussed below), and the infinite-order extrapolation is done only in Part~II.}

The
Feynman rules for this scheme are as follows:
\be
\Sigma_\Nr^{(n)}[\GtN] = \Sigma_{\rm bold}^{(n)}[\GtN]\ \ \ {\rm for}\ n\leq \Nr;
\label{eq:Sigma_n<N}
\ee
while for $n\geq \Nr+1$,
$\Sigma_\Nr^{(n)}[\GtN]$ is the sum of all bare diagrams, built with $\GtN$ as free propagator and the bare interaction vertex corresponding to $S_{\rm int}$,
which do not contain any insertion of a subdiagram %belonging to the skeleton self-energy graphs of order $\leq \Nr$.
contributing to $\Sigma^{(n)}_{\rm bold}[\GtN]$ with $n\leq \Nr$.
Indeed, each such insertion is exactly compensated by the corresponding counterterm.
To derive Eq.~(\ref{eq:Sigma_n<N}), we will use the relation
\be
\Sigma_\Nr(\xi) {\  \hat{=}\ } \sum_{n=1}^\infty \Sigma_{\rm bold}^{(n)}[G_\Nr(\xi)]\, \xi^n\label{eq:Sig_xi_Sig_bold}
\ee
{ where $\hat{=}$ stands for equality in the sense of formal power series in $\xi$,}
and we will show the proposition
\begin{equation*}
  \ \ \ \ \ \ \ \ \ \ \ \ \ \ \ \ \ \ \
  \Sigma_\Nr(\xi) {\  \hat{=}\ } \sum_{n=1}^k \Sigma_{\rm bold}^{(n)}[\GtN]\,\xi^n + O(\xi^{k+1})
\ \ \ \ \ \ \ \ \ \ \ (\Pr_k)
\end{equation*}
for any $k\in \{0,\ldots,\Nr+1\}$, by recursion over $k$.
$(\Pr_{k=0})$ clearly holds.
If $(\Pr_k)$ holds for some $k\leq\Nr$, then we have
$G_\Nr(\xi) {\  \hat{=}\ } \GtN + O(\xi^{k+1})$,
{ as follows from Eqs.~(\ref{eq:Dyson_N_xi}), (\ref{eq:G0_N_xi}) and (\ref{eq:Lambda_semi_b}).}
Substitution into Eq.~(\ref{eq:Sig_xi_Sig_bold}) then yields
$(\Pr_{k+1})$.

%for each $n\leq {\cal N}$,  the first $n$ counter-terms, $\Lambda_1, \Lambda_2, \ldots, \Lambda_n$,  exactly compensate all the  up-to-the-$n$-th order contributions to the self-energy coming from $S_{\rm int}$.
%
%The resulting expansion---standard for effective field theories---is then identical to the semi-skeleton series based on Dyson summation of infinite sets of irreducible diagrams associated with the first ${\cal N}$ orders of the perturbative expansion of the original action (\ref{action_phys}).

%In view of exact cancelation of all contributions up to order ${\cal N}$,
%the counter-terms $\Lambda_n$ do not enter the diagrammatic expansion explicitly;
%the only relevant entity is $\tilde{G}_{\cal N}$.

%A separate question is how to find  $ \tilde{G}_{\cal N}$ for a given $G_0$.
%But that is precisely what the method of bold diagrammatic Monte Carlo (BDMC) has been designed for.
%We thus arrive at the numeric protocol consisting of

  %For technical reasons,
Alternatively to 
%Note that one is not limited to
the semi-bold scheme Eq.~(\ref{eq:Lambda_semi_b}),
%since
other choices  are possible for the shifts $\Lambda_1, \ldots, \Lambda_\Nr$ and the dressed propagator $\GtN$.
  %one might prefer to work with a
%  $S_\Nr^{(\xi)}[\psi,\bar{\psi}]$ in which (some of) the shifts $\Lambda_n$ are different from the $n$-th order skeleton-type contributions to the
%self-energy originating from $S_{\rm int}$, e.g.
For example, the shifts can be based on diagrams containing
the original bare propagator $G_0$ %, or some other function,
%not
instead of
$ \tilde{G}_{\cal N}$. %, to avoid self-consistency in Eq.~(\ref{G_0_constr}) and guarantee the existence of its solution.
%In this case,
In the absence of exact cancellation,
all diagrams should be simulated
in Part II of the numerical protocol,
and
$\Lambda_n$ will enter the theory explicitly.
This flexibility of choosing the form of $\Lambda_n$'s, along with the obvious option of
exploring different ${\cal N}$'s, provides a tool for controlling systematic errors coming from
truncation of the $\xi$-series
 \footnote{ 
The case where $\Nr=1$ and $\Lambda_1$ is a number is known as Screened Perturbation Theory in thermal $\phi^4$ theory;
its extension to gauge theories is known as Hard-Thermal-Loop Perturbation Theory~\cite{revue_qcd_resummation_hot}.
We thank E. Braaten
%, E.~Iancu and J.-P. Blaizot 
for pointing this out. }.
 %A safe way to proceed is to make sure that the results obtained with a number of different auxiliary actions are consistent with each other within the error bars.

%%%%%%%%%%%%%%%%%%%%%%%%%%%%%
%{\it Limit of the skeleton sequence.}
{\it Skeleton scheme.}
We turn to the { conventional}
scheme in which diagrams are built on the { {\it fully}} dressed single-particle propagator.
The corresponding numerical protocol is identical to Part I of the above one, with the additional step of extrapolating $\Nr$ to infinity, as done in~\cite{ProkofevSvistunovPolaronLong,VanHouckeEOS,KulaginPRL,MishchenkoProkofevPRL2014,DengEmergentBCS}.
Accordingly, we assume that the `skeleton sequence' $\GtN$ converges to a limit $\Gt$ when $\Nr\to\infty$.
The crucial
question is under what conditions one can be confident that $\tilde{G}$ is the genuine propagator $G$ of the original model.
The answer comes from
the %convergence
properties of the sequence of functions
\be
%\Sigma_{\rm bold}^{(\leq\Nr)}[\GtN,\xi]
\bL_\Nr^{(\xi)} \coloneqq \sum_{n=1}^\Nr \Sigma_{\rm bold}^{(n)}[\GtN]\,\xi^n.
\ee
Let us show that $\Gt=G$ holds under the following sufficient condition:
\\(i) for any $\xi$ in a disc $\Dr=\{|\xi|<R\}$ of radius $R>1$,
and for all $(\pp, \tau)$,
%$\Sigma_{\rm bold}^{(\leq\Nr)}[\GtN,\xi](\pp,\tau)$
$\bL_\Nr^{(\xi)}(\pp,\tau)$
converges for $\Nr{\to}\infty$;
moreover this sequence is uniformly bounded, {\it i.e.},
there exists a function $C_1(\pp,\tau)$ such that $\forall \xi\in\Dr, \forall (\Nr,\pp,\tau)$,
%$|\Sigma_{\rm bold}^{(\leq\Nr)}[\GtN,\xi](\pp,\tau)|
$|\bL_\Nr^{(\xi)}(\pp,\tau)|\leq C_1(\pp,\tau)$; and
\\(ii) $\GtN(\pp,\tau)$ is uniformly bounded, {\it i.e.},
there exists a constant $C_2$ such that for all $(\Nr, \pp, \tau)$, $|\GtN(\pp,\tau)| \leq C_2$. %converges for $\Nr\to\infty$ and is uni

Our derivation is based on the action
\be
S^{(\xi)}_\infty \coloneqq \lim_{\Nr\to\infty} S^{(\xi)}_\Nr.
\ee
Clearly,
\be
S^{(\xi)}_\infty= \la \psi |\, \Gt^{-1} + \bL^{(\xi)} \, | \psi\ra + \xi S_{\rm int}
\ee
with
\be
\bL^{(\xi)}(\pp,\tau) \coloneqq \lim_{\Nr\to\infty} \bL_\Nr^{(\xi)}(\pp,\tau). %\Sigma_{\rm bold}^{(\leq\Nr)}[\GtN,\xi](\pp,\tau).
\ee
Since $S_\Nr^{(\xi=1)}=S$, we have $S_\infty^{(\xi=1)}=S$, and thus
$G_\infty(\xi{=}1)=G$ where $G_\infty(\xi)$ is the full propagator of the action $S^{(\xi)}_\infty$.

We first observe that
$\bL^{(\xi)}(\pp,\tau)$ is an analytic function of $\xi\in\Dr$ for all $(\pp,\tau)$,
and that
\be
 \frac{1}{n!} \left. \frac{\partial^n}{\partial \xi^n}\bL^{(\xi)}(\pp,\tau)\right|_{\xi=0}
  %= n!\,\lim_{\Nr\to\infty}\Sigma_{\rm bold}^{(n)}[\GtN](\pp,\tau)
 =\  \Sigma_{\rm bold}^{(n)}[\Gt](\pp,\tau).
 \label{eq:deriv_n}
\ee
This follows from conditions (i,ii), given that momenta are bounded for lattice models.
  Indeed, for any triangle $\mathcal{T}$ included in $\Dr$, $\oint_{\mathcal{T}} d\xi \,\bL_\Nr^{(\xi)}(\pp,\tau) =0$.
  %and the limit $\Nr\to\infty$ can be taken under the integral
  Thanks to condition (i), the dominated convergence theorem is applicable,
  yielding
 $\oint_{\mathcal{T}} d\xi \,\bL^{(\xi)}(\pp,\tau) =0$.
  The analyticity of $\xi\mapsto\bL^{(\xi)}(\pp,\tau)$ follows by Morera's theorem.
To derive  Eq.~(\ref{eq:deriv_n}) we start from
\be
\frac{1}{n!} \left.\frac{\partial^n}{\partial \xi^n} \bL^{(\xi)}_\Nr(\pp,\tau)\right|_{\xi=0}
=\  \Sigma_{\rm bold}^{(n)}[\GtN](\pp,\tau).
\label{eq:deriv_Nr}
\ee
By Cauchy's integral formula,
the l.h.s. of Eq.~(\ref{eq:deriv_Nr})  equals
$1/(2 i \pi) \oint_\Cr d\xi\, \bL^{(\xi)}_\Nr(\pp,\tau) / \xi^{n+1}$
where $\Cr$ is the unit circle.
Using again condition (i) and the dominated convergence theorem, when $\Nr{\to}\infty$, this tends to 
$1/(2 i \pi) \oint_\Cr d\xi\, \bL^{(\xi)}(\pp,\tau) / \xi^{n+1}$,
which equals the l.h.s. of Eq.~(\ref{eq:deriv_n}).
To show that $\Sigma_{\rm bold}^{(n)}[\GtN](\pp,\tau)$ tends to 
$\Sigma_{\rm bold}^{(n)}[\Gt](\pp,\tau)$,
we consider each Feynman diagram separately;
the dominated convergence theorem is applicable
%by taking  the limit $\Nr\to\infty$ on the l.h.s. using Cauchy's integral formula,  condition (i) and dominated convergence, and on the r.h.s. using dominated convergence thanks to
thanks to condition (ii), the boundedness of the integration domain for internal momenta and imaginary times,
and assuming that interactions decay sufficiently quickly at large distances for the bare interaction vertex to be bounded in momentum representation.

Hence
\be
\bL^{(\xi)}
= \sum_{n=1}^\infty \Sigma_{\rm bold}^{(n)}[\Gt]\ \xi^n.
\ee
As a consequence, the action $S_\infty^{(\xi)}$ generates the fully dressed skeleton series built on $\Gt$,
{\it i.e.},
its self-energy $\Sigma_\infty(\xi)$ has the Taylor expansion
$\sum_{n=1}^\infty \Sigma_{\rm bold}^{(n)}[\Gt]\,\xi^n$,
and the Taylor series of $G_\infty(\xi)$ reduces to the $\xi$-independent term $\Gt$.
This can be derived in the same way as Eq.~(\ref{eq:Sigma_n<N}),
{ by showing by recursion over $k$ that for any $k\geq0$,
$  \Sigma_\infty(\xi) = \sum_{n=1}^k \Sigma_{\rm bold}^{(n)}[\Gt]\,\xi^n + O(\xi^{k+1})$.}
Furthermore,
having shown
{
above  the analiticity %at $\xi=0$,
of $L^{(\xi)}$, {\it i.e.}, of
$S^{(\xi)}_\infty$,} % is analytic 
we again expect that
$G_\infty(\xi)$ is analytic at $\xi=0$
(for fermions on a lattice at finite temperature),
%Besides, from the analyticity of $S^{(\xi)}_\infty$ for $\xi\in\Dr$,
%we infer that $G_\infty(\xi)$ is also analytic for $\xi\in\Dr$~[\thefnnumber].
and we can use again the above presumption to conclude that
$G_\infty(\xi=1) = G = \Gt$.

%%%%%%%%%%%%%%%%%%%%%%%%%%%%%

{\it Dressed pair propagator.}
So far we have discussed dressing of the single-particle propagator while keeping the bare interaction vertices.
We turn to diagrammatic schemes built on dressed pair propagators.
%Restricting for simplicity to spin-$1/2$ fermions with a contact interaction, the interaction part of the action writes
We restrict %for simplicity
to spin-$1/2$ fermions with on-site interaction:
\begin{multline}
  S_{\rm int}[\psi,\bar{\psi}] = U\, \sum_{\rr} \int_0^\beta d\tau\ 
(\bar{\psi}_\up \bar{\psi}_\down \psi_\down \psi_\up)(\rr,\tau),
\end{multline}
where $U$ is the bare interaction strength.
For simplicity we discuss dressing of the pair propagator while keeping the bare $G_0$.
%For notational transparency, we will discuss dressing of the pair propagator only; constructing counter-terms for both single-particle and pair-propagator is nothing but a combination of these two cases.
It is necessary to perform a Hubbard-Stratonovich transformation in order to construct the appropriate auxiliary action.
Introducing a complex scalar Hubbard-Stratonovich field $\eta$ leads to the action
\begin{multline}
\Sr[\psi,\bar{\psi},\eta,\bar{\eta}] =   \la\psi|G_0^{-1}|\psi\ra - \la\eta|\Gamma_0^{-1}|\eta\ra
-  %\xi
\la\eta|\Pi_0|\eta\ra
  \\
  +%\sqrt{\xi}
  \,
  \la\eta|\psi_\down\psi_\up\ra
  +
  \la\psi_\down\psi_\up|\eta\ra,
\end{multline}
where $\Pi_0(\rr,\tau) = - (G_{0,\up} G_{0,\down})(\rr,\tau)$
and $\Gamma_0$ is the sum of the ladder diagrams,
$\Gamma_0^{-1}(\pp,\Omega_n) = U^{-1} - \Pi_0(\pp,\Omega_n)$
with $\Omega_n$ the bosonic Matsubara frequencies.

We first consider the diagrammatic scheme built on $G_0$ and $\Gamma_0$.
%(in which $\Gamma_0$  a partially dressed interaction vertex
We denote by $\Sigma^{(n)}_{\rm lad}[G_0,\Gamma_0]$ the sum of all self-energy diagrams of order $n$, {\it i.e.} containing $n$ $\Gamma_0$-lines.
This diagrammatic series is generated by the shifted action
\begin{multline}
\Sr_{\rm lad}^{(\xi)}[\psi,\bar{\psi},\eta,\bar{\eta}] =   \la\psi|G_0^{-1}|\psi\ra - \la\eta|\Gamma_0^{-1}|\eta\ra
-  \xi^2
\la\eta|\Pi_0|\eta\ra
  \\
  +\xi
  \,\big(
  \la\eta|\psi_\down\psi_\up\ra
  +
  \la\psi_\down\psi_\up|\eta\ra
  \big),
\end{multline}
in the sense that self-energy $\Sigma_{\rm lad}(\xi)$ corresponding to this action has the Taylor series
$\sum_{n=1}^\infty \Sigma^{(n)}_{\rm lad}[G_0,\Gamma_0]\, \xi^{2n}$.
Indeed, the counter-term $\xi^2 \Pi_0$ cancels out the reducible diagrams contatining $G_0 G_0$ bubbles.
Therefore, if this diagrammatic series converges, then it yields the physical self-energy. This follows from the same reasoning as below Eq.~(\ref{eq:Dyson_N_xi}).
The same applies to the series for the pair self-energy $\Pi$ in terms of $[G_0,\Gamma_0]$. Here $\Pi$ is defined by
$\Gamma^{-1} = \Gamma_0^{-1} - \Pi$,
where $\Gamma$ denotes the fully dressed pair propagator, used in~\cite{VanHouckeEOS,DengEmergentBCS}.

More complex schemes,
built on other dressed pair propagators than $\Gamma_0$,
can be generated by the shifted action 
%The pair propagator can be dressed further %analogously to the above single-particle propagator case The corresponding auxiliary action reads
\begin{multline}
  \Sr_{\Nr}^{(\xi)}[\psi,\bar{\psi},\eta,\bar{\eta}] =   \la\psi|G_0^{-1}|\psi\ra - \la\eta|\Gamma_{0,\Nr}^{-1}(\xi)|\eta\ra
  -  \xi^2
\la\eta|\Pi_0|\eta\ra
  \\
  +\xi
  \,\big(
  \la\eta|\psi_\down\psi_\up\ra
  +
  \la\psi_\down\psi_\up|\eta\ra
  \big),
\end{multline}
where
\be
\Gamma_{0,\Nr}^{-1}(\xi) = \tilde{\Gamma}_\Nr^{-1} + \xi^2\,\Omega_1 + \ldots + \xi^{2\Nr} \, \Omega_\Nr
\ee
and one imposes $\Gamma_{0,\Nr}(\xi=1) = \Gamma_0$.
In particular, the semi-bold scheme is defined by
\be
\Omega_n = \Pi_{\rm bold}^{(n)}[\tilde{\Gamma}_\Nr],
\ee
where $\Pi_{\rm bold}^{(n)}[\gamma]$ is the sum of all skeleton diagrams of order $n$ built with the pair-propagator $\gamma$ that remain connected when cutting two $\gamma$-lines. As usual, $\Pi_{\rm bold}^{(1)} = - GG + G_0 G_0$.
This scheme was introduced previously for $\Nr{=}1$~\cite{KrisUnpub}.

Finally, we consider the skeleton scheme built on $G_0$ and $\Gamma$.
Assuming that the skeleton sequence $\tilde{\Gamma}_\Nr$ converges to some $\tilde{\Gamma}$, one can show analogously to the above reasoning
that $\tilde{\Gamma}$ is equal to the exact $\Gamma$ under the following sufficient condition:
\\(i) for any $\xi$ in a disc $\Dr=\{|\xi|<R\}$ of radius $R>1$,
and for all $(\pp, \Omega_n)$,
$M_\Nr^{(\xi)}(\pp,\Omega_n) \coloneqq \sum_{n=1}^\Nr \Pi_{\rm bold}^{(n)}[\tilde{\Gamma}_\Nr](\pp, \Omega_n)\,\xi^n$
converges for $\Nr{\to}\infty$;
moreover this sequence
is uniformly bounded, {\it i.e.},
there exists $C(\pp,\Omega_n)$ such that $\forall \xi\in\Dr, \forall(\Nr,\pp,\Omega_n), |M_\Nr^{(\xi)}(\pp,\Omega_n)| \leq C(\pp,\Omega_n)$;
%and for all $(\pp, \Omega_n)$,
%$\Sigma_{\rm bold}^{(\leq\Nr)}[\GtN,\xi](\pp,\tau)$
%$\sum_{n=1}^\Nr \Pi_{\rm bold}^{(n)}[\tilde{\Gamma}_\Nr]\,\xi^n$
%converges for $\Nr{\to}\infty$
%and
%is uniformly bounded with respect to $\xi$ and $\Nr$;
and
\\(ii) $\tilde{\Gamma}_\Nr(\pp,\Omega_n)$ is uniformly bounded.

%%%%%%%%%%%%%%%%%%%%%%%%%%%%%%%

{\it Screened interaction potential.}
%As an intermediate step between the above-discussed case of dressing the single-particle propagator and the general (semi-)skeleton technique, consider
Finally, we briefly address
the procedure of dressing the interaction line,
which is particularly important for long-range interaction potentials.
Restricting for simplicity to a spin-independent interaction potential $V(\rr)$, the interaction part of the action writes
\be
  %  S_{\rm int}[\psi,\bar{\psi}] = \\
  \frac{1}{2}\sum_{\sigma,\sigma'} \sum_{
    \rr, \rr'} \int_0^\beta d\tau\ 
(\bar{\psi}_\sigma \psi_\sigma)(\rr,\tau)
\,V(\rr-\rr')\,
(\bar{\psi}_{\sigma'} \psi_{\sigma'})(\rr',\tau).
\ee
We again keep the bare $G_0$ for simplicity and
consider dressing of $V$ only.
%; constructing counter-terms for both $G$ and $V$ is nothing but a combination of these two cases.
%It is necessary to perform a Hubbard-Stratonovich transformation in order to construct the appropriate auxiliary action.
 Introducing a real scalar Hubbard-Stratonovich field $\chi$ leads to the action
%the pairwise potential interaction $S_{\rm int}[\psi,\bar{\psi}]$ decomposes as follows ($S_{\rm int}[\psi,\bar{\psi}]  \, \to \, {\cal S}[\psi,\bar{\psi}, \chi]$):
\begin{equation}
  \Sr[\psi,\bar{\psi}, \chi]\, =\, \la\psi|G_0^{-1}|\psi\ra + {1\over 2}\langle \chi | \, V^{-1} \, | \chi \rangle  \, +\,
    i\,
    \sum_\sigma \la \chi | \bar{\psi}_\sigma \psi_\sigma \ra .
\label{HS}
\end{equation}
%    Here the bra-ket notation implies summation over space and imaginary time,
Here
we assume 
that the Fourier transform $V(\qq)$ of the interaction potential is positive,
so that the quadratic form
$\langle \chi | \, V^{-1} \, | \chi \rangle = (2\pi)^{-d}  \int_0^\beta
d\tau \int d^d q \,
|\chi(\qq,\tau)|^2 / V(\qq) $
is positive definite.
The 
auxiliary action %and the condition of correspondence have the form very close to that of (\ref{action_aux}) and (\ref{G_0_constr}), respectively:
takes the form
\begin{multline}
  \Sr^{(\xi)}_\Nr[\psi,\bar{\psi},\chi]  =
  \la\psi|G_0^{-1}|\psi\ra
  \\+
  \frac{1}{2}
\langle \chi | \tilde{V}_{\cal N}^{-1} \! + \xi^2 \Omega_1   + \ldots + \xi^{2{\cal N}} \Omega_{\cal N}| \chi \rangle
   %- \la\psi|G_0^{-1}|\psi\ra  \,
+\,  i \xi\, \sum_\sigma \la \chi | \bar{\psi}_\sigma \psi_\sigma \ra . %\qquad \qquad \qquad
%  \nonumber
\label{action_aux_2}
\end{multline}
%and the condition $\Sr^{(\xi=1)}_\Nr = \Sr$ writes
%\begin{equation}
% \tilde{V}_{\cal N}^{-1} + \Omega_1 + \Omega_2   + \ldots + \Omega_{\cal N} \, =\,  V^{-1}.
%\label{V_constr}
%\end{equation}
The semi-bold scheme corresponds to $\Omega_n = \Pi_{\rm bold}^{(n)}[\tilde{V}_\Nr]$ where $\Pi$ now stands for the polarization. In particular, $\tilde{V}_1$ is the RPA screened interaction.

%%%%%%%%%%%%%%%%%%%%%%%%%%%%%

Summarizing, we have revealed an analytic structure behind dressed-line diagrammatics.
More precisely, we have exhibited the function which analytically continues a dressed diagrammatic series.
This function originates
from an action that depends on an auxiliary parameter $\xi$.
%that plays the role of coupling constant in a shifted action $S_\Nr^{(\xi)}$, provided $S_{\xi =1}^{({\cal N})} $ coincides with the physical action $S$. The expansion is made on top of harmonic action  $S^{(\xi = 0)}_\Nr$ and, formally, takes the form of standard perturbative expansion in $\xi$-dependent interaction and $\xi$-dependent harmonic counter-terms (shifts).
When the action is a polynomial in $\xi$, the situation reduces to the one of a bare expansion.
%Particularly simple Feynman rules are obtained in
Within this category,
a particular case
well suited for numerical implementation
is
the semi-bold scheme
%expansion is obtained when
for which
the bare propagator %and shifts in  $S_\Nr^{(\xi)}$ are
is taken from the truncated bold self-consistent equation. % truncated self-consistent formulation.
%This scheme is applicable even when the fully bold scheme is not.
For the fully bold scheme,
we construct an appropriate auxiliary action,
but only under a certain condition.
If this condition is verified numerically, it is safe to use the fully bold scheme.
%This provides a criterion for the validity of the fully bold scheme.
If not, the semi-bold scheme remains applicable.

Furthermore we have demonstrated the generality of the shifted-action construction by treating the case of a dressed pair propagator and of a screened long-range interaction. %({\it e.g.} built on ladder diagrams)
Further extensions left for future work are dressing of three-point vertices, as well as
justifying resummation of divergent diagrammatic series { by considering non disc-shaped analyticity domains $\Dr$}.

We are grateful to Youjin Deng, Evgeny Kozik, Kris Van Houcke, for valuable exchanges.
This work was supported by the National Science Foundation under the grant PHY-1314735,
The Simons Collaboration on the Many Electron Problem,
the MURI Program ``New Quantum Phases of Matter" from AFOSR,
CNRS, ERC and IFRAF-NanoK
(grants PICS 06220, {\it Thermodynamix},
and
{\it Atomix}).
R.R. and F.W. are affiliated to {\it Paris~Sciences~et~Lettres, Sorbonne~Universit\'es}, and {\it Sorbonne~Paris~Cit\'e}.

\bibliography{felix_copy}

\end{document}